# Photonic spin Hall effect dependent on Landau level transitions in monolayer $WTe_2$

Qiaoyun Ma,[1] Hui Dou,[1] Yiting Chen,[2] Guangyi Jia,[1,*] and Xinxing Zhou[2,3,4,†]

[1]*School of Science, Tianjin University of Commerce, Tianjin 300134, P.R. China*

[2]*Key Laboratory of Low-Dimensional Quantum Structures and Quantum Control of Ministry of Education, School of Physics and Electronics, Hunan Normal University, Changsha 410081, P.R. China*

[3]*Key Laboratory of Physics and Devices in Post-Moore Era, College of Hunan Province, Changsha 410081, P.R. China*

[4]*Institute of Interdisciplinary Studies, Hunan Normal University, Changsha 410081, P.R. China*

[*]gyjia87@163.com; [†]xinxingzhou@hunnu.edu.cn

Landau level (LL) engineered photonic spin Hall effect (PSHE) holds great promise for nanoscale manipulation and steering of magneto-optical transport in two-dimensional atomic systems. Herein, we theoretically investigate PSHE modulated by LL transitions $\delta n = n' - n = 0, \pm 2$ (where $n$ and $n'$ indicate the LL indexes of valence and conduction bands, respectively) in monolayer $WTe_2$. Results show that PSHE tuned by $\delta n = 0, \pm 2$ has completely different dependent behaviors on LLs, incident angle of incident photons, and magnetic induction intensity. These discrepancies are ascribed to Hall-conductivity-incurred Hall angle $\Theta$ because the variation tendency of photonic spin Hall shifts is similar to that of $\Theta$ with changing the LL index. Giant PSHE with the largest in-plane displacement of more than 400 times of incident wavelength is obtained at the transition $|n=55\rangle \rightarrow |n'=57\rangle$. Remarkably enhanced PSHE occurs at near-zero Hall angles. In-plane and transverse spin-dependent displacements give their respective extremum values at the same incident angles when the $\Theta$ is near to zero, and their incident-angle deviation will become larger and larger as the $|\Theta|$ increases. This unambiguously confirms the strong influence of Hall angle in the PSHE, shedding important insights into the fundamental properties of spin-orbit interaction of light in time-reversal symmetry breaking quantum systems.

## I. INTRODUCTION

The response of two-dimensional (2D) electron gases to an external magnetic field empowers 2D materials with many intriguing magneto-optical transport merits [1,2]. One of the most representative merits is the transformation of continuous electronic energy band into a series of discrete Landau levels (LLs). LLs are cornerstones of the quantum Hall effect, where the quantization of Hall conductance is directly related to the filling of LLs. And the energy separation between adjacent LLs is determined by the magnetic field strength and LL index. Furthermore, band parameters including the Fermi velocity and the effective mass of charge carriers are retrievable from the experimentally acquired LL spectrum [3,4]. This analytical approach has already demonstrated its effectiveness in characterizing a range of materials, such as graphene, black phosphorus, and topological metals, etc [4-8]. By virtue of these admirable properties, LL splitting opens up new avenues in the manipulation of photonic spin Hall effect (PSHE), which can not only quantizedly enhance the spin splitting magnitude of photons but also realize precise tuning of the PSHE response via external magnetic field modulation-allowing dynamic control over the spin-dependent lateral shift of light beams without altering material's self-structure [9-11].

PSHE is generally accepted as a result of the longitudinal spin-orbit interaction, which describes the mutual influence of the longitudinal spin (polarization) and trajectory of light beam [12-14]. It is also regarded as a direct photonic analog of the electronic spin Hall effect, in which the electron spin and electric potential gradient are substituted by the optical helicity of incident photons and the refractive index gradient, respectively. In conventional optical processes, PSHE occurs at the sub-wavelength scale due to the inherent weak spin-orbit interaction. In recent years, researchers have attempted to enhance the PSHE by imposing a tunable magnetic field [9-11], and the synergistic combination of matter magnetization and PSHE manipulation has shown great promise for cutting-edge applications including barcode encryption [11], probing quantum geometric capacitance [15], and sensing deep-subwavelength disorders in materials [16], and so forth. Magnetic-field-modulated PSHE has been studies in many materials, however, most of them rely on intrinsic magnetic ordering or magnetic dopants to induce magneto-optical responses and do not involve LLs [13-18]. To our knowledge, LL-engineered PSHE has been only reported in graphene, black phosphorus, and topological silicene thus far [9-11,19,20]. More unfortunately, these scarce studies are mainly focused on superficial quantization phenomenon-with little attention paid to the variation of Hall angle induced by LL transitions-and lack systematic investigations on the correlation between LL index, Hall angle evolution, and photonic spin splitting [9-11,19,20]. The in-depth regulatory mechanism of PSHE by different LL transitions remains ambiguous.

In the present work, we perform a theoretical study on LL-dependent PSHE in anisotropic monolayer $WTe_2$. The model based on an effective ***k·p*** Hamiltonian and linear-response theory is built to get LL spectra and magneto-optical conductivity tensor of $WTe_2$ for PSHE analyses. We explore how LL transitions $\delta n = 0, \pm 2$, LL index $n$ ranging from 1 to 61, and LL-dependent Hall angle impact in-plane and transverse photonic spin Hall shifts. Results show that giant spin Hall shifts, hundreds of times of the incident wavelength, can be obtained at appropriate LL transitions. The possible regulatory mechanism is discussed in detail.

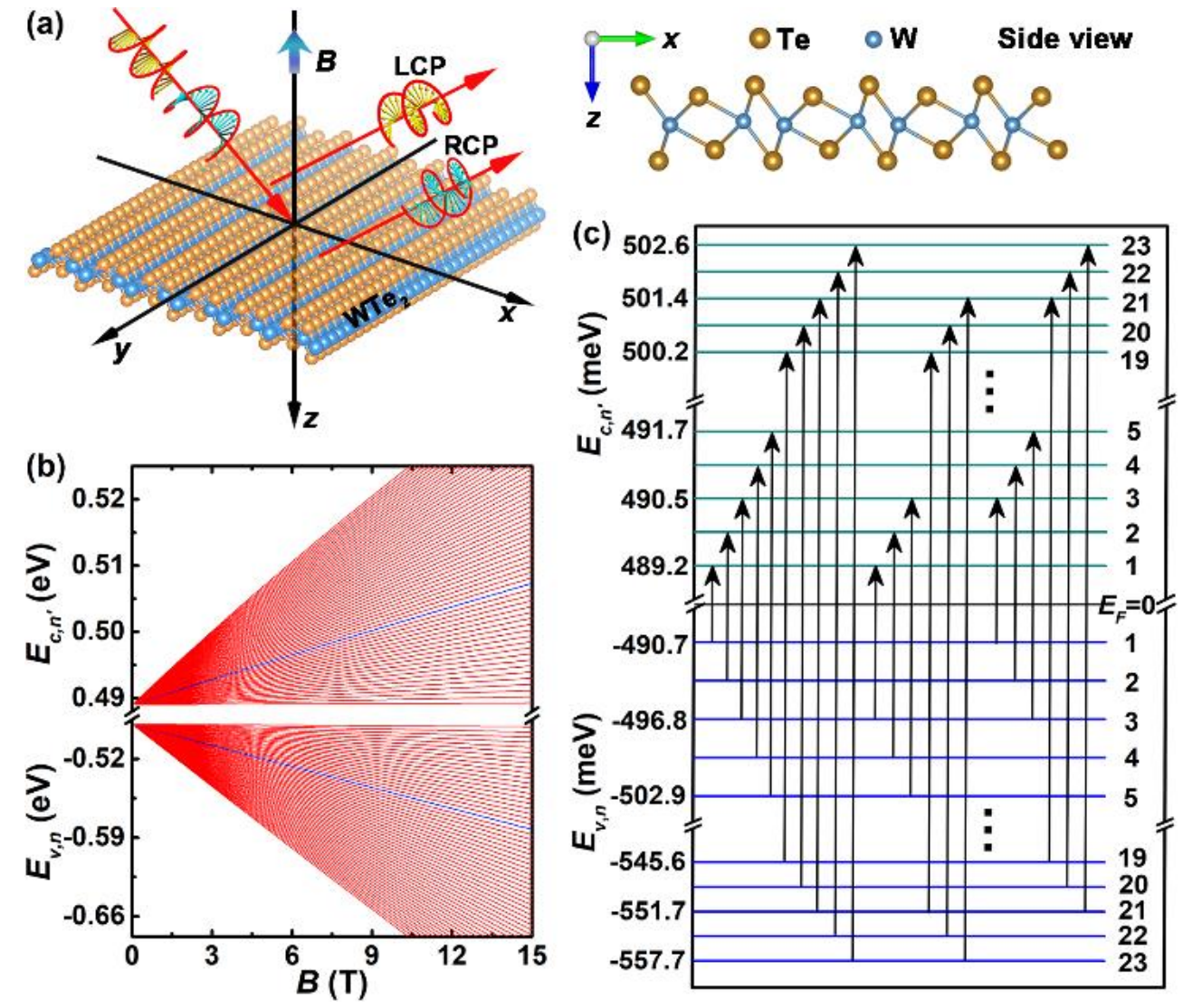

FIG. 1. (a) Schematic diagram of the photonic spin splitting of one Gaussian beam reflected from free-suspended monolayer $WTe_2$. An imposed static magnetic field ***B*** is applied along the negative $z$ axis in a Cartesian coordinate system. Upper right subgraph: side view of the crystallographic structure of $WTe_2$. (b) LLs of conduction and valence bands of monolayer $WTe_2$ in different magnetic fields. LLs at $n = n' = 21$ are marked by blue lines. The number of basis function used in the calculation is 200 to get convergent numerical results. (c) Schematic illustration of the interband transition rules at $B$ = 10 T, where numbers on right side indicate LL indexes (LLs $6 \leq n(n') \leq 18$ are not sketched and are implied by ellipses).

## II. MODEL AND THEORETICAL METHODS

In monolayer 1T′-$WTe_2$, the valence band mainly consists of $d$-orbitals of W atom, while the conduction band mainly consists of $p_y$-orbitals of Te atom. Its low energy ***k·p*** Hamiltonian can be expressed as [21-23]

$$H_{k\cdot p}=\begin{pmatrix} E_c\left(k_x,k_y\right) & 0 & -i\gamma_1 k_x & \gamma_2 k_y \\ 0 & E_c\left(k_x,k_y\right) & \gamma_2 k_y & -i\gamma_1 k_x \\ i\gamma_1 k_x & \gamma_2 k_y & E_v\left(k_x,k_y\right) & 0 \\ \gamma_2 k_y & i\gamma_1 k_x & 0 & E_v\left(k_x,k_y\right) \end{pmatrix}, \qquad (1)$$

where $E_c\left(k_x,k_y\right)=-\Delta-\alpha k_x^2-\beta k_y^2$ and $E_v\left(k_x,k_y\right)=\Delta+\lambda k_x^2+\eta k_y^2$ are the on-site energies of $p$ and $d$ orbitals in $WTe_2$, respectively, with $\Delta = -0.489$ eV representing the $d$-$p$ band inversion at $\Gamma$ point in the Brillouin zone. The factors $\alpha$, $\beta$, $\lambda$, and $\eta$ are related to the reduced Planck's constant $\hbar$ and effective masses by $\alpha=\hbar^2/(2m_{cx})$, $\beta=\hbar^2/(2m_{cy})$, $\lambda=\hbar^2/(2m_{vx})$, and $\eta=\hbar^2/(2m_{vy})$, respectively. The parameters $\gamma_1=\hbar v_1$ and $\gamma_2=\hbar v_2$ with $v_1$ and $v_2$ denoting the velocities along the $x$ and $y$ directions, respectively. These factors were obtained by fitting the first-principles band structures, and they are $\alpha$=11.25 eV·Å$^2$, $\beta$=6.90 eV·Å$^2$, $\lambda$=0.27 eV·Å$^2$, $\eta$=1.08 eV·Å$^2$, $\gamma_1$=1.71 eV·Å, and $\gamma_2$=0.48 eV·Å [23].

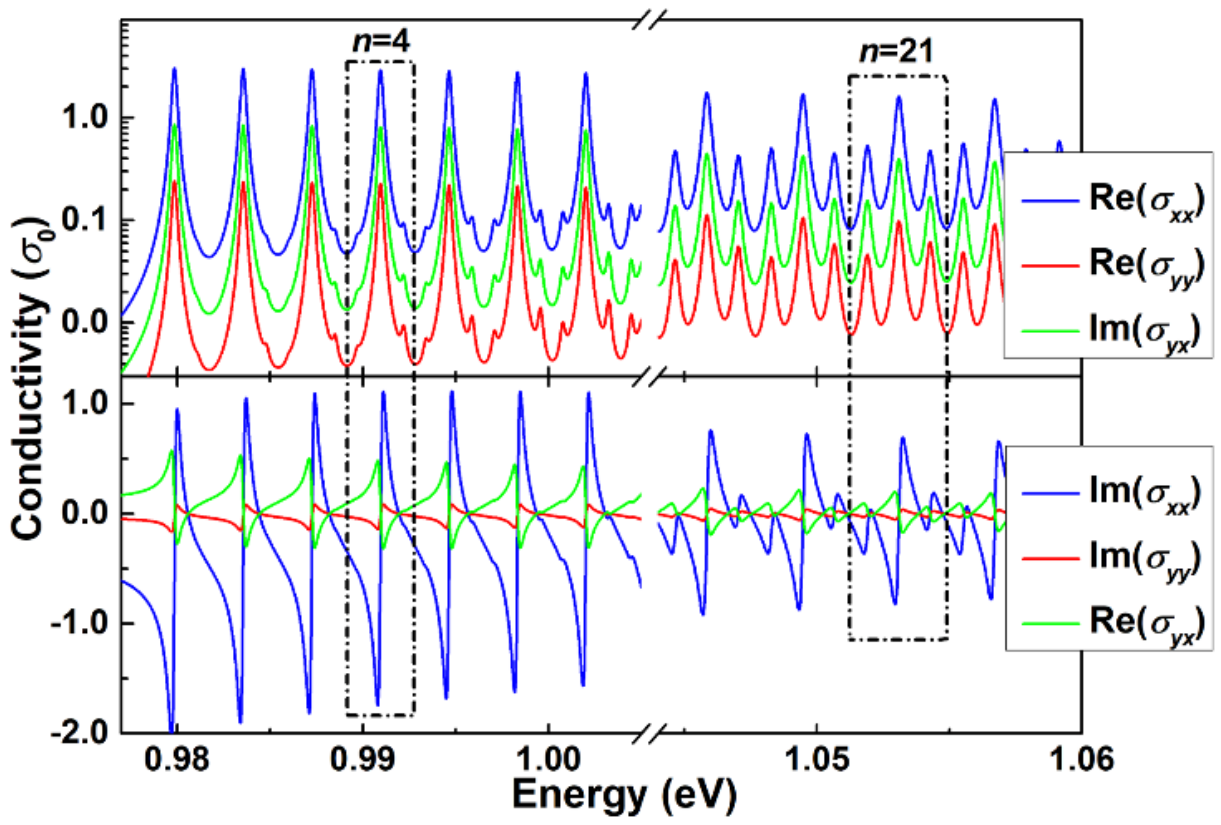

FIG. 2. Optical conductivity spectra of monolayer $WTe_2$ versus the $\hbar\omega$ values at $B$ = 10 T. Oscillation periods at LL indexes $n$ = 4 and 21 are marked by two dashed-line boxes.

On the basis of a unitary transformation [22,23], the 4×4 Hamiltonian $H_{k\cdot p}$ in Eq. (1) can be separated into two decoupled 2×2 Hamiltonian blocks corresponding to "spin-up" and "spin-down" states, respectively. These two 2×2 blocks share the same dispersion relation. Thus, for brevity, we can only consider the spin-up Hamiltonian that reads [22,23]

$$H=\begin{pmatrix} E_c\left(k_x,k_y\right) & -iv_1P_x-v_2P_y \\ iv_1P_x-v_2P_y & E_v\left(k_x,k_y\right) \end{pmatrix}, \tag{2}$$

where $P_x = \hbar k_x$ and $P_y = \hbar k_y$.

To describe the PSHE in a general model, monolayer $WTe_2$ is assumed to be suspended in a free space where both the permittivities $\varepsilon_1$ and $\varepsilon_2$ of upper and bottom media are 1.0. One Gaussian wave packet with monochromatic frequency $\omega$ impinges upon the surface of $WTe_2$ film with an incident angle $\theta_i$. The $x$ and $y$ axes of laboratory Cartesian coordinate ($x$, $y$, $z$) are along the armchair and zigzag orientations of $WTe_2$, respectively, as sketched in Fig. 1(a). And a uniform static magnetic field $\boldsymbol{B} = -B\hat{\boldsymbol{z}}$ is applied along the negative $z$ axis. Within the linear-response theory [8,10], the magneto-optical response of monolayer $WTe_2$ is characterized by an optical conductivity tensor

$$\hat{\sigma}(\omega)=\begin{pmatrix} \sigma_{xx} & \sigma_{xy} \\ \sigma_{yx} & \sigma_{yy} \end{pmatrix}, \tag{3}$$

with

$$\sigma_{jk}=i\sigma_0\frac{\hbar^2}{l_B^2}\sum_{n',n,c,v}\frac{\left[f(E_{c,n'})-f(E_{v,n})\right]\langle c,n'|v_j|v,n\rangle\langle v,n|v_k|c,n'\rangle}{(E_{c,n'}-E_{v,n})(E_{c,n'}-E_{v,n}+\hbar\omega+i\Gamma)}, \tag{4}$$

where $j$, $k\in\{x, y\}$, $\sigma_{xy} = -\sigma_{yx}$ is Hall conductivity, $n$ (or $n'$) is the LL index of valence (or conduction) band, $\sigma_0=e^2/\pi\hbar$, $l_B^2=\hbar/(eB)$, $f(E_\xi)=\{\exp[(E_\zeta - E_F)/k_BT]+1\}^{-1}$ is the Fermi-Dirac distribution function with Boltzman constant $k_B$ and temperature $T$; $E_{v,n}$ and $E_{c,n'}$ are the LLs of valence and conduction bands, respectively; and $v_{j/k}$ are the components of group velocities. The sum runs over all states $|\zeta\rangle= |v,n\rangle$ and $|\zeta'\rangle = |c,n'\rangle$ with $\zeta \neq \zeta'$. In this work, the level broadening factor $\Gamma$ and temperature $T$ are set to be 0.15 meV and 5 K, respectively. Besides, we take the Fermi energy $E_F$=0 such that the contribution from intraband transitions can be ignored [10,24]. Fig. 1(b) shows the LLs as a function of the magnetic field for the first 61 LLs in monolayer $WTe_2$. More detailed introductions on calculations of LLs and transition matrix elements of the velocity matrices $v_{j/k}$ can be found from Eqs. (S1)-(S8) in [25].

According to the generalized 4×4 transfer-matrix formalism [26], the Fresnel reflection coefficients of free-suspended monolayer $WTe_2$ are calculated by

$$r_{pp}=-\frac{P_{12}^{--}S_{12}^{++}+\sigma_{xy}\sigma_{yx}}{P_{12}^{++}S_{12}^{++}-\sigma_{xy}\sigma_{yx}}, \tag{5}$$

$$r_{sp}=-\frac{2\sigma_{xy}}{P_{12}^{++}S_{12}^{++}-\sigma_{xy}\sigma_{yx}}, \tag{6}$$

$$r_{ps}=\frac{2\sigma_{yx}}{P_{12}^{++}S_{12}^{++}-\sigma_{xy}\sigma_{yx}}, \tag{7}$$

$$r_{ss}=\frac{P_{12}^{++}S_{12}^{--}+\sigma_{xy}\sigma_{yx}}{P_{12}^{++}S_{12}^{++}-\sigma_{xy}\sigma_{yx}}, \tag{8}$$

with $P_{12}^{\pm\pm}=1/\cos\theta_i\pm 1/\cos\theta_i\pm\sigma_{xx}$ and $S_{12}^{\pm\pm}=\cos\theta_i\pm\cos\theta_i\pm\sigma_{yy}$. In the above Eqs. (5)-(8), all the conductivity tensor components are normalized to $c\varepsilon_0/4\pi$ where $c$ and $\varepsilon_0$ stand for light velocity in vacuum and vacuum permittivity, respectively.

PSHE manifests itself as spin-dependent splitting, which appears in both transverse and in-plane directions. In the spin basis set, the linearly polarized state of light wave can be decomposed into two orthogonal spin components, i.e., the left- and right-circular polarization (LCP and RCP) components [10,27]. Under the paraxial approximation, photonic spin Hall shifts of these two orthogonal components in the reflected beam can be mathematically derived [27,28]. PSHE is sensitive to the polarization states of incident light. For simplicity, only $p$-polarized incident beam is considered in this paper. The corresponding in-plane and transverse spin Hall shifts for LCP and RCP components, as indicated in Fig. 1(a), are derived from [10,27]

$$\Delta x_{|\pm\rangle} = \mp \frac{1}{k_0} \mathrm{Re}\left( \frac{r_{pp}}{r_{pp}^2 + r_{sp}^2}\frac{\partial r_{sp}}{\partial\theta_i} - \frac{r_{sp}}{r_{pp}^2+r_{sp}^2}\frac{\partial r_{pp}}{\partial\theta_i}\right), \tag{9}$$

$$\Delta y_{|\pm\rangle} = \mp\frac{\cot\theta_i}{k_0}\mathrm{Re}\left(\frac{r_{pp}+r_{ss}}{r_{pp}^2+r_{sp}^2} r_{pp} - \frac{r_{ps}-r_{sp}}{r_{pp}^2+r_{sp}^2} r_{sp}\right), \tag{10}$$

respectively. Here, $k_0 = 2\pi/\lambda_0$ is the wavevector in vacuum ($\lambda_0 = 2\pi c/\omega$ indicates the excitation wavelength), the superscripts $|+\rangle$ and $|-\rangle$ imply the LCP and RCP components, respectively. It is seen from the above equations that the inter-LL transitions can be used to modify the PSHE in virtue of their decisive impact on the conductivity tensor. Because $\Delta x_{|+\rangle} = -\Delta x_{|-\rangle}$ and $\Delta y_{|+\rangle} = -\Delta y_{|-\rangle}$, we will only analyze the photonic spin Hall shifts for the LCP component in the following.

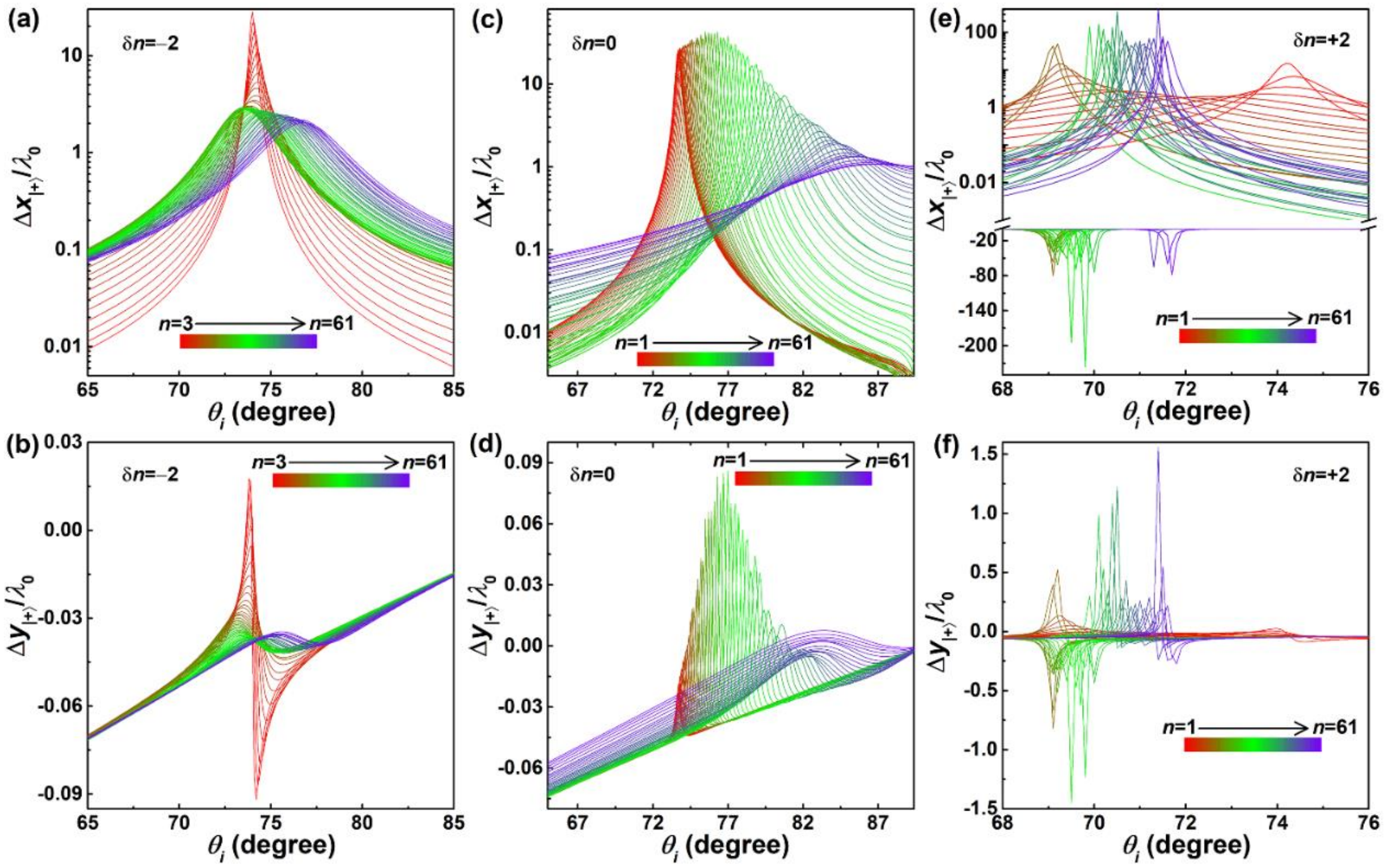

FIG. 3. (a)(c)(e) In-plane and (b)(d)(f) transverse spin-dependent shifts on the surface of monolayer $WTe_2$ versus the incident angle $\theta_i$ and LL index $n$. (a)(b), (c)(d), and (e)(f) are for LL transitions $\delta n = n' - n = -2$, 0, and +2, respectively. The magnetic field $B = 10$ T.

## III. RESULTS AND DISCUSSIONS

Fig. 2 presents partially magnified $\sigma_{xx}$, $\sigma_{yy}$ and $\sigma_{yx}$ spectra of monolayer $WTe_2$ at $B = 10$ T. Their overall spectra ranging the $\hbar\omega$ from 0.98 to 1.20 eV are given in Figs. S1-S3 in Ref. [25]. One can see that all the conductivity spectra are quantized and oscillated with increasing the photonic energy, and a three-peak structure is well resolved in each oscillation period in the real or imaginary part of $\sigma_{xx}$ ($\sigma_{yy}$) or $\sigma_{yx}$. The magneto-optical transition selection rules are determined by the transition matrix elements of velocity matrices in Eqs. (S7) and (S8) in Ref. [25]. By matching with LLs in Fig. 1(b), these three peaks from left to right in $\mathrm{Re}(\sigma_{xx,yy})$ correspond to interband transitions when the LL index changes $\delta n = n' - n = -2$, 0, and +2. In Fig. 2, oscillation periods at $n = 4$ and 21 are marked by two dashed-line boxes, and interband transition rules are sketched in Fig. 1(c).

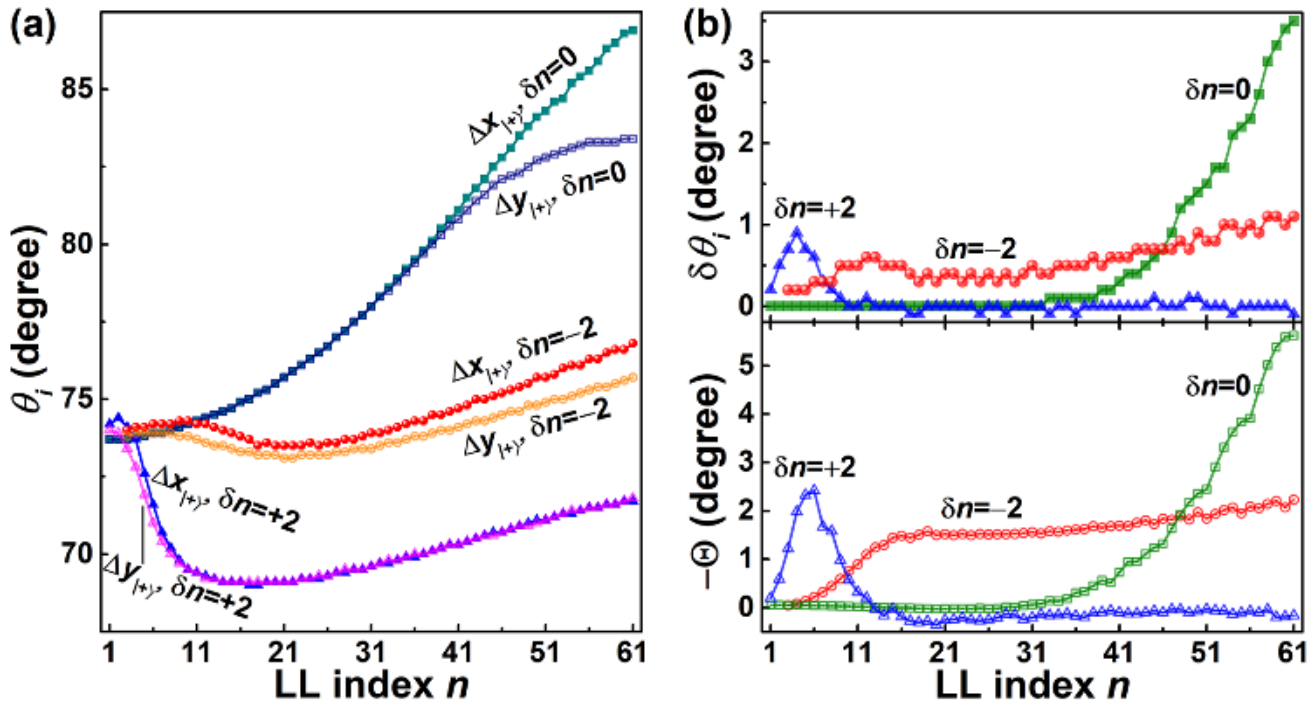

FIG. 4. (a) LL-dependent incident angles at which $\Delta x_{|+\rangle}$ and $\Delta y_{|+\rangle}$ have extremum displacements. (b) Upper subgraph: angle differences between incident angles of extremum displacements of $\Delta x_{|+\rangle}$ and $\Delta y_{|+\rangle}$. Bottom subgraph: variations of the Hall angle −Θ with respect to LLs. Symbols are the calculated values while solid lines are just drawn as a guide to the eye.

After substituting these LL-dependent conductivity components in Figs. 2 and S1-S3 into Eqs. (5)-(10), the optical spectra of photonic spin Hall shifts on the surface of monolayer $WTe_2$ can be derived. Fig. 3 shows the variations of in-plane and transverse spin-dependent shifts with the incident angle $\theta_i$ at increasing $n$ from 1 (or 3) to 61 for $\delta n$ = 0, +2 (or $\delta n$ = −2). To more clearly distinguish the curve shape, Fig. S4 illustrates the optical spectra of $\Delta x_{|+\rangle}$ and $\Delta y_{|+\rangle}$ at inter-LL transitions $|n=3\rangle \rightarrow |n'=1\rangle$, $|n=21\rangle \rightarrow |n'=21\rangle$, and $|n=55\rangle \rightarrow |n'=57\rangle$. It is seen that the shift spectrum for each inter-LL transition gives a peak (it exhibits a dip for negative shifts). These peak or dip values correspond to their respective extremum displacements. Fig. 4(a) summarizes the $\theta_i$ values of these extremum displacements. The extremum displacements of $\Delta x_{|+\rangle}$ have a similar evolvement rule with that of $\Delta y_{|+\rangle}$ at tunning LL index $n$, and the former is far larger than the latter. Thus, LL-dependent extremum displacements of $\Delta x_{|+\rangle}$ and $\Delta y_{|+\rangle}$ are summarized in Figs. 5 and S5, respectively. Besides, Figs. 5 and S5 also show the variations of conductivity differences $\mathrm{Im}(\sigma_{xx})-\mathrm{Im}(\sigma_{yy})$ and $\mathrm{Im}(\sigma_{xx})-\mathrm{Re}(\sigma_{yx})$ with LL index $n$. We find that the jagged spectra of conductivity differences are accompanied with slight quantum oscillations in the spectra of photonic spin Hall shifts.

For the case of $\delta n$ = 0, it is seen from Fig. 4(a) that the peak position of $\Delta x_{|+\rangle}$ (or $\Delta y_{|+\rangle}$ ) gradually increases from 73.7° to 86.9°(or 83.4°) as the LL index $n$ increases from 1 to 61. Extremum displacements of $\Delta x_{|+\rangle}$ and $\Delta y_{|+\rangle}$ appear at the same $\theta_i$ value at $n \leq 32$, but they deviate from each other at $n > 32$ and their angle difference $\delta\theta_i$ becomes larger and larger as the $n$ increases [see upper subgraph in Fig. 4(b)]. Even if the slight quantum oscillation, the main variation trend of extremum displacements in Fig. 5(c) is that the displacements firstly increase and then decrease with an increment of the $n$. Especially, the in-plane spin-dependent shift $\Delta x_{|+\rangle}$ gives the largest value of 42.496$\lambda_0$ at $n$ = 21 [see Fig. 5(c)].

For the cases of $\delta n$ = ±2 [*cf.* Figs. 3(a)(b) and (e)(f)], their photonic spin Hall shifts have quite different LL-dependent behaviors, both of which are also different from that of $\delta n$ = 0. For example, the in-plane spin-dependent shift $\Delta x_{|+\rangle}$ for $\delta n$ = −2 has the largest displacement of 28.037$\lambda_0$ at $n$ = 3 ($n'$ = 1), and it dramatically decreases to 3.028$\lambda_0$ as the $n$ increases to 15 [see Fig. 5(b)]. By contrast, the $\Delta x_{|+\rangle}$ for $\delta n$ = +2 gives the largest value of 401.022$\lambda_0$ at $n$ = 55, and its extremum displacements are strikingly oscillated with changing the $n$ [see Fig. 5(d)]. The $\theta_i$ values of extremum displacements for $\delta n$ = 0, ±2 also show completely different LL-dependent behaviors, as demonstrated in Fig. 4(a).

The variations of photonic spin Hall shifts with LLs could be attributed to LL-dependent Hall angle Θ. Hall angle describes the efficiency of converting a longitudinal driving current into a lateral Hall current, which is calculated via $\Theta = \arctan(\sigma_{yx}/\sigma_{xx})$ [29-31]. Fig. 4(b) shows that the variation tendency of angle difference $\delta\theta_i$ is similar to that of −Θ with changing the $n$. Extremum displacements of $\Delta x_{|+\rangle}$ and $\Delta y_{|+\rangle}$ appear at the same incident angle (i.e., $\delta\theta_i$ = 0) when the Θ is near to zero. If −Θ is large enough, the angle difference $\delta\theta_i \neq 0$ and it mainly increases with the increasing of −Θ. This means that the incident-angle alignment of in-plane and transverse spin Hall shifts is governed by the magnitude of |Θ|. As the |Θ| increases, the anisotropy introduced by Hall conductivity breaks this alignment, resulting in a growing $\delta\theta_i$. This

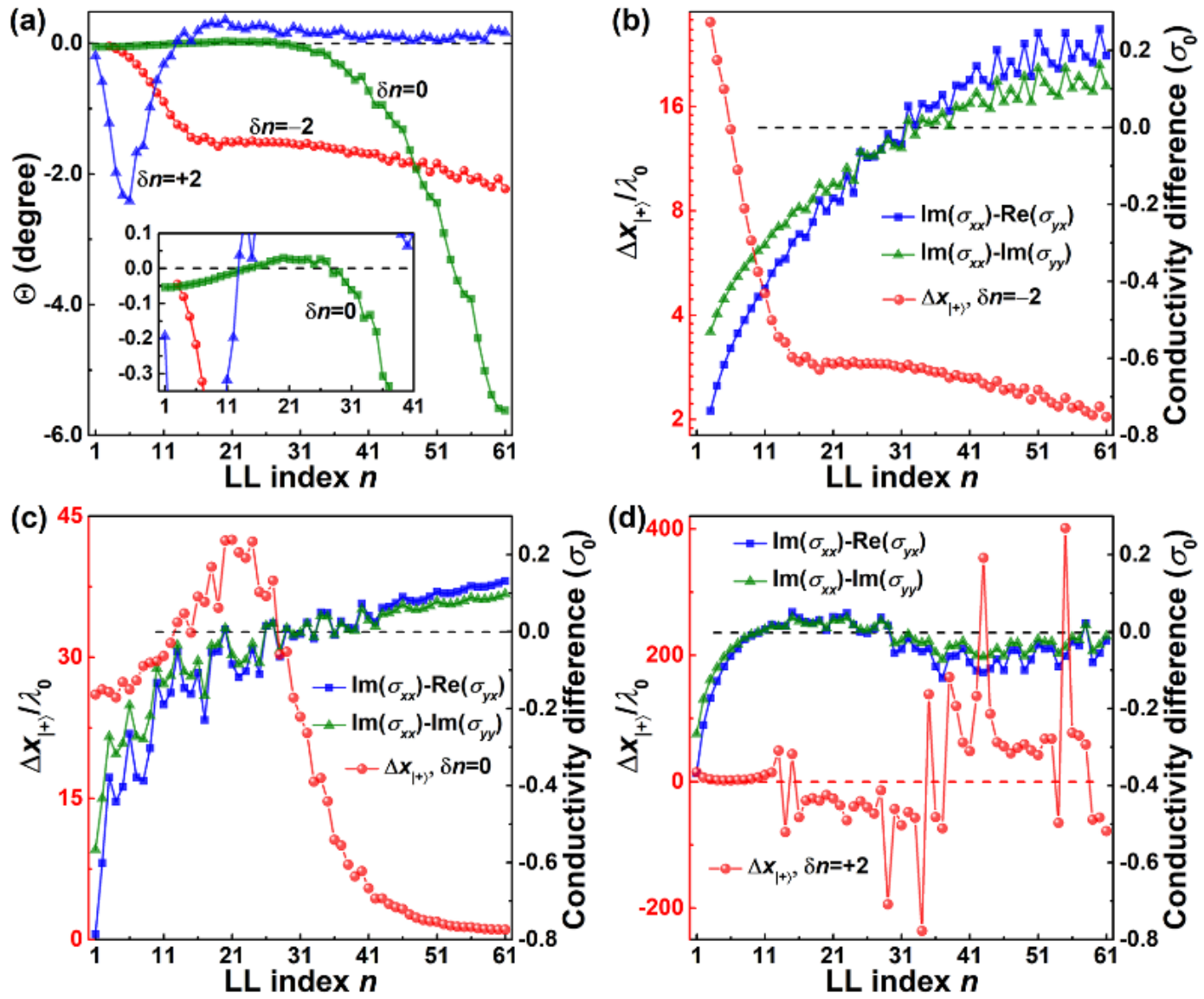

FIG. 5. (a) Variations of the Hall angle Θ with respect to LLs. The inset shows partially magnified Θ spectra for clarity. (b)-(d) Evaluations of the largest in-plane spin-dependent shifts with LL index $n$. (b), (c), and (d) are for the inter-LL transitions $\delta n = -2, 0, +2$, respectively. For comparison, LL-dependent conductivity differences $\mathrm{Im}(\sigma_{xx})-\mathrm{Im}(\sigma_{yy})$ and $\mathrm{Im}(\sigma_{xx})-\mathrm{Re}(\sigma_{yx})$ are also shown in (b)-(d). The graduation lines at which the vertical coordinate axis equals zero are marked by dashed horizontal lines. Symbols are the calculated values while solid lines are just drawn as a guide to the eye. The magnetic field $B = 10$ T.

directly links the macroscopic PSHE behavior to the microscopic Hall transport property.

Fig. 5 shows that extremum displacements of PSHE are also impacted by Hall angle. For instance, at $\delta n = 0$, LL-dependent Θ spectrum in Fig. 5(a) exhibits a wide peak around $n = 21$, matching with the peak position ($n = 21$) of LL-dependent $\Delta x_{|+\rangle}$ spectrum in Fig. 5(c). At $\delta n = -2$, the Θ is near to zero at $n = 3$ ($n' = 1$) and sharply decreases to $-1.44°$ as the $n$ increases to 15, being in line with the curve shape of $\Delta x_{|+\rangle}$ spectrum in Fig. 5(b). At $\delta n = +2$, both the Θ and $\Delta x_{|+\rangle}$ spectra in the range $1 \le n \le 12$ exhibits a dip [*cf.* Figs. 5(a) and (d)], and the largest displacement of $\Delta x_{|+\rangle}$ is only $14.927\lambda_0$ in this range. In contrast, the Θ oscillates near 0° at $n \ge 13$ such that the $\Delta x_{|+\rangle}$ also strikingly oscillates and gives relatively larger displacements (the largest displacement $401.022\lambda_0$ appears at $n = 55$ with $\Theta = 0.11°$).

According to Figs. 5 and S5, we construct an empirical formula between photonic spin Hall shifts with Hall angle by

$$F(\Theta) = \sum_{i=1}^{5} a_i \exp\left[-\left(\frac{\Theta - b_i}{c_i}\right)^2\right]. \tag{11}$$

In the above equation, $F(\Theta)$ can be either the in-plane or transverse spin Hall shift, the factors $a_i$, $b_i$, and $c_i$ are coefficients and their detailed values are listed in Table S1 [25]. Fig. S6 presents the extremum displacements of $\Delta x_{|+\rangle}$ and $\Delta y_{|+\rangle}$ as a function of Hall angle Θ. It is found that $F(\Theta)$ is in accordance with the variation tendencies of $\Delta x_{|+\rangle}$ and $\Delta y_{|+\rangle}$ calculated via Eqs. (9) and (10). This empirical relationship provides a quantitative tool to predict PSHE magnitude from Hall angle measurements. The physical picture underlying Eq. (11) is rooted in the interplay between Hall conductivity and spin-orbit interaction of light. It captures an important trend: PSHE is a non-linear function of Hall angle, peaking at near-zero Θ and decaying exponentially as |Θ| deviates from zero. The coefficients $a_i$, $b_i$, and $c_i$ are derived from numerical calculations, validating the universality of this trend across different LL transitions ($\delta n = 0, \pm 2$).

On the basis of the above discussions, it is concluded that the LL dependence of PSHE is a direct manifestation of LL-tunable Hall angle. Large |Θ| reduces the magnitudes of photonic spin splitting and breaks the alignment of incident angles at which $\Delta x_{|+\rangle}$ and $\Delta y_{|+\rangle}$ have extremum displacements. By contrast, near-zero Θ is a "sweet spot" for PSHE enhancement, at which the spin-orbit interaction of light is strong and $\delta\theta_i = 0$. Even so, Hall angle Θ can not be strictly zero if we want to obtain a giant PSHE. This is because that the Hall conductivity $\sigma_{xy} = 0$ will lead to $r_{sp} = r_{ps} = 0$ according to Eqs. (6) and (7). In consequence, the $\Delta x_{|+\rangle}$ will equal to zero and the $\Delta y_{|+\rangle}$ will become far smaller than excitation wavelength $\lambda_0$ based on Eqs. (9) and (10). These ideas are further evidenced by modulating the magnetic induction intensity $B$.

Figs. 6(a) and S7 show the changes of complex optical conductivities of monolayer $WTe_2$ with the $B$ values at LL index $n = 21$. Accordingly, magnetic-field-dependent Hall angle and photonic spin Hall shifts can be derived, as shown in Figs. 6(b) and (c), respectively. For LL transitions $\delta n$ = −2 and 0, both of their Hall angles are gradually decreased with increasing the $B$ value. Thereby, extremum displacements of $\Delta x_{|+\rangle}$ and $\Delta y_{|+\rangle}$ are also gradually decreased with enhancing the magnetic field. Because the Θ values of $\delta n$ = −2 are far away from 0°, the $\Delta x_{|+\rangle}$ (or $\Delta y_{|+\rangle}$) of $\delta n$ = −2 is far smaller than that of $\delta n$ = 0 [*cf.* red and green symbols in Fig. 6(c)]. For the LL transition $\delta n$ = +2, the optical spectra of conductivity components Im($\sigma_{xx}$), Im($\sigma_{yy}$), and Re($\sigma_{xy}$) cross around the coordinate (0$\sigma_0$, 8 T), as shown by blue symbols in Fig. 6(a). As a result, photonic spin Hall shifts $\Delta x_{|+\rangle}$ and $\Delta y_{|+\rangle}$ of $\delta n$ = +2 in Fig. 6(c) change from positive to negative values as the $B$ increases from 8 T to be ≥ 9 T. As the $B$ increases from 5 T to 8 T, the Θ increases from −0.06° to 0.12°, correspondingly, the extremum displacement of $\Delta x_{|+\rangle}$ (or $\Delta y_{|+\rangle}$) gradually increases from 25.817$\lambda_0$ to 271.683$\lambda_0$ (or from 0.003$\lambda_0$ to 1.002$\lambda_0$). At $B$ ≥ 9 T, the Θ gradually deviates from 0° as the $B$ increases, thus, the in-plane and transverse displacements (i.e., $|\Delta x_{|+\rangle}|$ and $|\Delta y_{|+\rangle}|$) gradually decrease with enhancing the magnetic field.

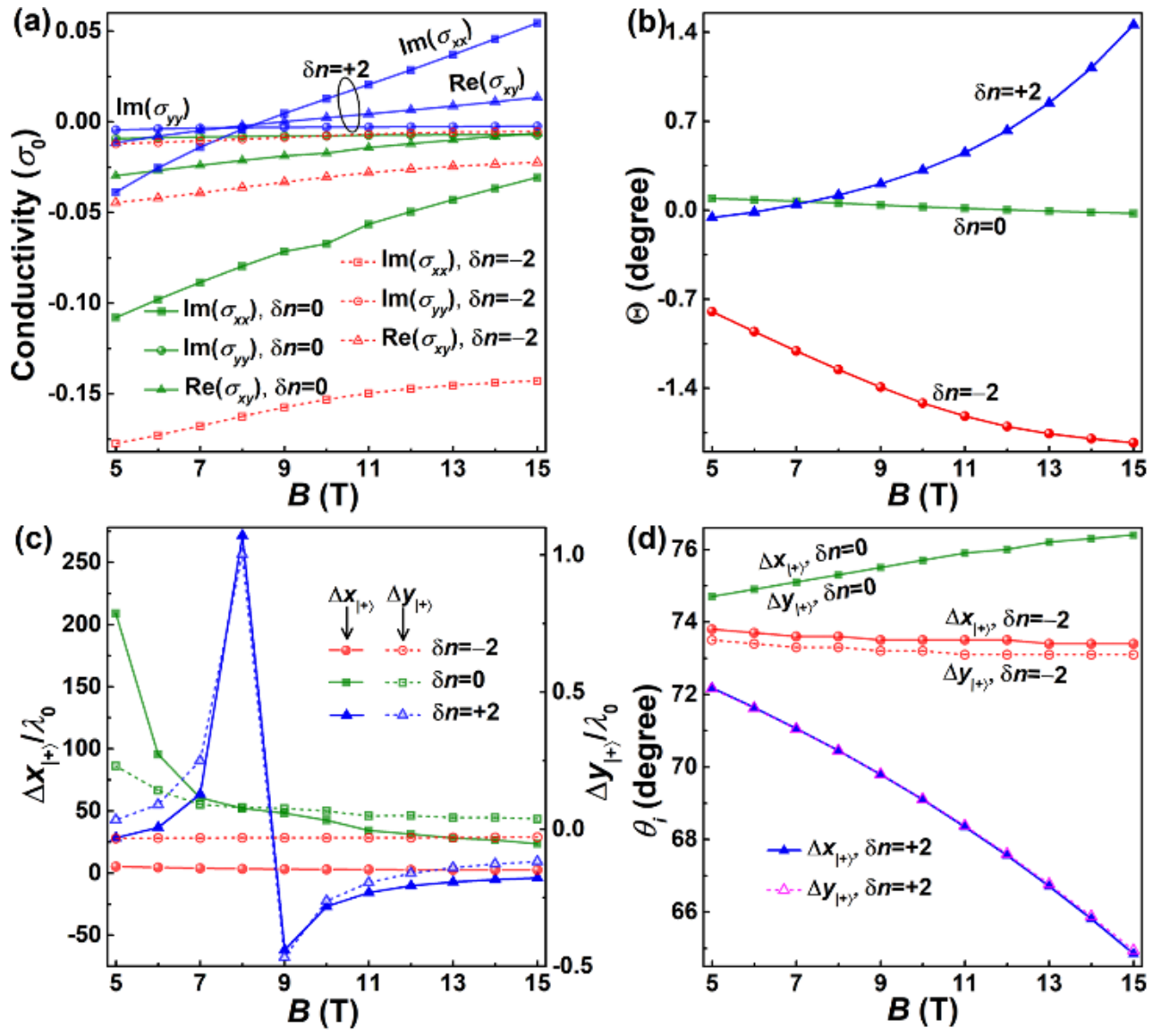

FIG. 6. At the transitions $\delta n$ = 0, ±2 with $n$ = 21, (a) complex optical conductivities versus the magnetic induction intensity $B$; (b) variations of the Hall angle Θ with respect to the $B$; (c) extremum displacements of $\Delta x_{|+\rangle}$ (solid symbols, refer to left vertical axis) and $\Delta y_{|+\rangle}$ (open symbols, refer to right vertical axis) versus the $B$; (d) magnetic-field-dependent incident angles at which $\Delta x_{|+\rangle}$ and $\Delta y_{|+\rangle}$ have the extremum displacements. Symbols are the calculated values while solid or dashed lines are just drawn as a guide to the eye.

Fig. 6(d) exhibits magnetic-field-dependent $\theta_i$ values at which $\Delta x_{|+\rangle}$ and $\Delta y_{|+\rangle}$ have extremum displacements. Because Hall angles are close to 0° (or far away from 0°) at $\delta n$ = 0 (or $\delta n$ = −2), $\Delta x_{|+\rangle}$ and $\Delta y_{|+\rangle}$ present their respective extremum displacements at the same (or different) incident angles, as marked by green (or red) symbols in Fig. 6(d). For the transition $\delta n$ = +2, extremum displacements of $\Delta x_{|+\rangle}$ and $\Delta y_{|+\rangle}$ appear at nearly the same incident angle at $B$ ≤ 12 T. When the magnetic field is $B$ ≥ 13 T, Hall angles of $\delta n$ = +2 become larger than 0.7° such that the incident angle of extremum displacement of $\Delta y_{|+\rangle}$ begins to deviate from that of $\Delta x_{|+\rangle}$, as marked by open and solid triangle symbols in Fig. 6(d).

Finally, we also calculate the photonic spin Hall shifts on the surface of monolayer $WTe_2$ without the external magnetic field. At this condition, Hall conductivity $\sigma_{xy}$ equals to zero, longitudinal and transverse optical conductivities of monolayer $WTe_2$ are calculated by the Lorentz-Drude model [32]. Fig. S8 shows the optical conductivity spectra of $WTe_2$ at $B$ = 0 T. Corresponding photonic spin Hall shifts, as a function of the photonic energy and the incident angle, are shown in Fig. S9. Results show that all the photonic spin Hall shifts are smaller than 0.15$\lambda_0$. Therefore, a non-vanishing Hall conductivity together with near-zero Hall angle is responsible for the aforementioned giant PSHE.

## IV. CONCLUSIONS

To summarize, LL-engineered PSHE has been studied in monolayer $WTe_2$. Results show that PSHE induced via LL transitions $\delta n$ = 0, ±2 in $WTe_2$ exhibit completely different dependent behaviors on LL index. For $\delta n$ = 0, the extremum displacements of PSHE mainly increase with increasing LL index up to $n$ = 21 and decreases thereafter. For $\delta n$ = −2, photonic spin Hall shifts decrease sharply as $n$ increases to 15, and then tend to slow variations. For $\delta n = +2$, the extremum displacement is strikingly oscillated with changing the $n$ and it gives the largest value (more than 400 times of the incident wavelength) at $n = 55$. PSHE induced via LL transitions $\delta n = 0, \pm 2$ also present different physical responses at tuning the incident angle of incident photons and magnetic induction intensity. These discrepancies are attributed to Hall-conductivity-incurred Hall angle $\Theta$ because the variation tendency of extremum displacements of PSHE is similar to that of $\Theta$ with changing the LL index. Remarkably enhanced PSHE occurs at near-zero Hall angles. In-plane and transverse extremum displacements appear at different incident angles when the value of $|\Theta|$ is large enough, and their incident-angle deviation becomes larger and larger with the increasing of $-\Theta$. These results provide insight into the fundamental properties of spin-orbit interaction of light in time-reversal symmetry breaking 2D quantum systems. They may also open up new possibilities for integrating quantum Hall physics with spin-photon interactions, facilitating the development of novel quantum photonic devices.

## ACKNOWLEDGMENTS

The author G. Jia acknowledges the financial support by National Natural Science Foundation of China (Grant No. 11804251). The author X. Zhou acknowledges the financial support by National Natural Science Foundation of China (Grant Nos. 12374273, 12421005) and Hunan Provincial Major Sci-Tech Program (Grant No. 2023ZJ1010).

## DATA AVAILABILITY

The data that support the findings of this article are not publicly available. The data are available from the authors upon reasonable request.

Supplemental Material for

# Photonic spin Hall effect dependent on Landau level transitions in monolayer $WTe_2$

Qiaoyun Ma,[1] Hui Dou,[1] Yiting Chen,[2] Guangyi Jia,[1,*] and Xinxing Zhou[2,3,4,†]

[1]*School of Science, Tianjin University of Commerce, Tianjin 300134, P.R. China*
[2]*Key Laboratory of Low-Dimensional Quantum Structures and Quantum Control of Ministry of Education, School of Physics and Electronics, Hunan Normal University, Changsha 410081, P.R. China*
[3]*Key Laboratory of Physics and Devices in Post-Moore Era, College of Hunan Province, Changsha 410081, P.R. China*
[4]*Institute of Interdisciplinary Studies, Hunan Normal University, Changsha 410081, P.R. China*
[*]gyjia87@163.com; [†]xinxingzhou@hunnu.edu.cn

## I. Calculations on Landau levels and transition matrix elements

When a perpendicular magnetic field $\boldsymbol{B} = (0, 0, \boldsymbol{B})$ is applied in Fig. 1(a) in the main text, taking the Landau gauge $\boldsymbol{A} = (-By, 0, 0)$, the creation and annihilation operators can be defined as

$$\hat{a} = \sqrt{\frac{m_{cy}\omega_c}{2\hbar}}\left(y - y_0 + i\frac{P_y}{m_{cy}\omega_c}\right), \tag{S1}$$

$$\hat{a}^{\dagger} = \sqrt{\frac{m_{cy}\omega_c}{2\hbar}}\left(y - y_0 - i\frac{P_y}{m_{cy}\omega_c}\right), \tag{S2}$$

where $\omega_c = 2(\alpha\beta)^{1/2}/(\hbar l_B^2)$ is the cyclotron frequency, $y_0 = l_B^2 k_x$ is the cyclotron center, and $l_B^2 = \hbar/(eB)$ stands for the magnetic length. Then the Hamiltonian in Eq. (2) in the main text becomes

$$H = \begin{pmatrix} -\Delta - \hbar\omega_c\left(\hat{a}^{\dagger}\hat{a} + 1/2\right) & -i\hbar\omega_{dx}\left(\hat{a} + \hat{a}^{\dagger}\right) + i\hbar\omega_{dy}\left(\hat{a} - \hat{a}^{\dagger}\right) \\ i\hbar\omega_{dx}\left(\hat{a} + \hat{a}^{\dagger}\right) + i\hbar\omega_{dy}\left(\hat{a} - \hat{a}^{\dagger}\right) & \Delta + \hbar\omega_v\left(\hat{a}^{\dagger}\hat{a} + 1/2\right) + \hbar\omega'\left(\hat{a}^2 + \hat{a}^{\dagger 2}\right) \end{pmatrix}, \tag{S3}$$

where $\hbar\omega_{dx} = -\gamma_1(\beta/\alpha)^{1/4}/(\sqrt{2}\, l_B)$, $\hbar\omega_{dy} = \gamma_2(\alpha/\beta)^{1/4}/(\sqrt{2}\, l_B)$, $\omega_v = (r_x+r_y)\omega_c$ and $\omega' = (r_x-r_y)\omega_c/2$ with $r_x = \lambda/(2\alpha)$ and $r_y = \eta/(2\beta)$.

The eigenvalues and eigenvectors can be numerically evaluated by taking the eigenvectors of Hamiltonian as basis functions. In this basis, the wavefunction of the system reads [10,24]

$$\psi(x,y)=\frac{e^{ik_x x}}{\sqrt{L_x}}\sum_{m=0}^{M}\begin{pmatrix} c_m \\ d_m \end{pmatrix}\phi_m\left[\kappa(y-y_0)\right], \tag{S4}$$

where $\kappa=\sqrt{m_{cy}\omega_c/\hbar}$, $\phi_m(y)=e^{-y^2/2}H_m(y)/\sqrt{\sqrt{\pi}2^m m!}$ is the harmonic oscillator wave functions. Then, we can diagonalize the Hamiltonian in Eq. (S3) numerically in a truncated Hilbert space and obtain the eigenvalues as well as the eigenvectors. By using $v_{x/y}=\partial H/\partial P_{x/y}$, the velocity matrices can be derived as follows

$$v_x=\begin{pmatrix} f_1\left(\hat{a}+\hat{a}^\dagger\right) & -iv_1 \\ iv_1 & -2r_x f_1\left(\hat{a}+\hat{a}^\dagger\right) \end{pmatrix}, \tag{S5}$$

$$v_y=\begin{pmatrix} if_2\left(\hat{a}-\hat{a}^\dagger\right) & -v_2 \\ -v_2 & -2ir_y f_2\left(\hat{a}-\hat{a}^\dagger\right) \end{pmatrix}, \tag{S6}$$

where $f_1=[2\alpha(\alpha\beta)^{1/2}]^{1/2}/(\hbar l_B)$ and $f_2=[2\beta(\alpha\beta)^{1/2}]^{1/2}/(\hbar l_B)$. By means of the wave functions in Eq. (S4), the transition matrix elements of the velocity matrices for the same subband are calculated as

$$\begin{aligned}\left\langle c,n'\middle|v_x\middle|v,n\right\rangle=\sum_{m',m}^{M}\Big[&iv_1\left(d_{m'}^{n',c*}c_m^{n,v}-c_{m'}^{n',c*}d_m^{n,v}\right)\delta_{m',m}+f_1\sqrt{m}\left(c_{m'}^{n',c*}c_m^{n,v}-2r_x d_{m'}^{n',c*}d_m^{n,v}\right)\delta_{m',m-1}\\ &+f_1\sqrt{m+1}\left(c_{m'}^{n',c*}c_m^{n,v}-2r_x d_{m'}^{n',c*}d_m^{n,v}\right)\delta_{m',m+1}\Big]\end{aligned} \tag{S7}$$

$$\begin{aligned}\left\langle v,n\middle|v_y\middle|c,n'\right\rangle=\sum_{m',m}^{M}\Big[&-v_2\left(c_m^{n,v*}d_{m'}^{n',c}+d_m^{n,v*}c_{m'}^{n',c}\right)\delta_{m',m}+if_2\sqrt{m}\left(c_m^{n,v*}c_{m'}^{n',c}-2r_y d_m^{n,v*}d_{m'}^{n',c}\right)\delta_{m',m-1}\\ &-if_2\sqrt{m+1}\left(c_m^{n,v*}c_{m'}^{n',c}-2r_y d_m^{n,v*}d_{m'}^{n',c}\right)\delta_{m',m+1}\Big]\end{aligned} \tag{S8}$$

The total transition probability is the sum of all the subbands below the photon energy. Substituting Eqs. (S7) and (S8) into Eq. (4) in the main text, one can obtain the longitudinal (transverse) and Hall magneto-optical conductivities of monolayer $WTe_2$.

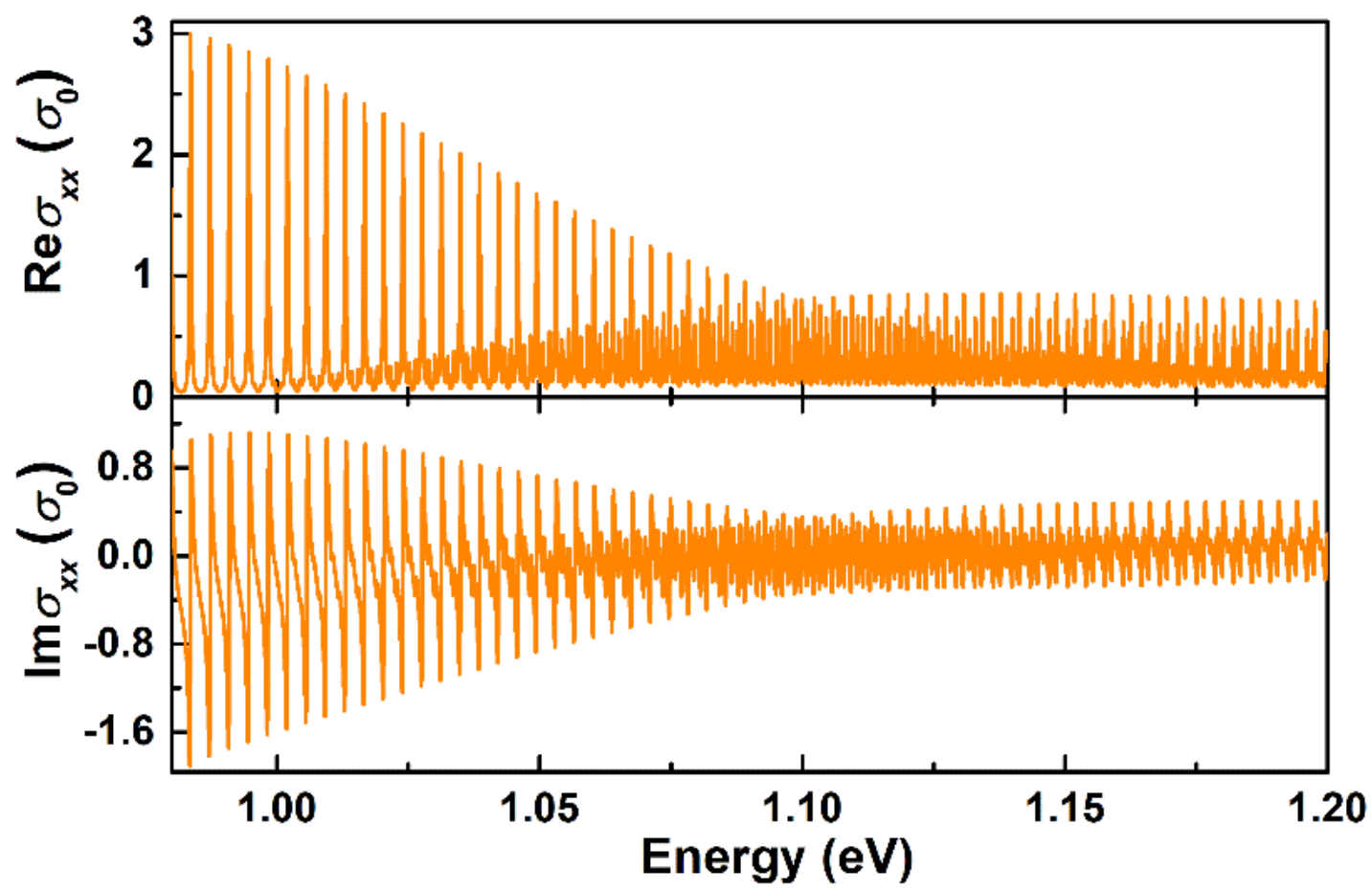

**Fig. S1** Real and imaginary parts of $\sigma_{xx}$ as a function of the photon energy for monolayer $WTe_2$ in a magnetic field with $B$ = 10 T.

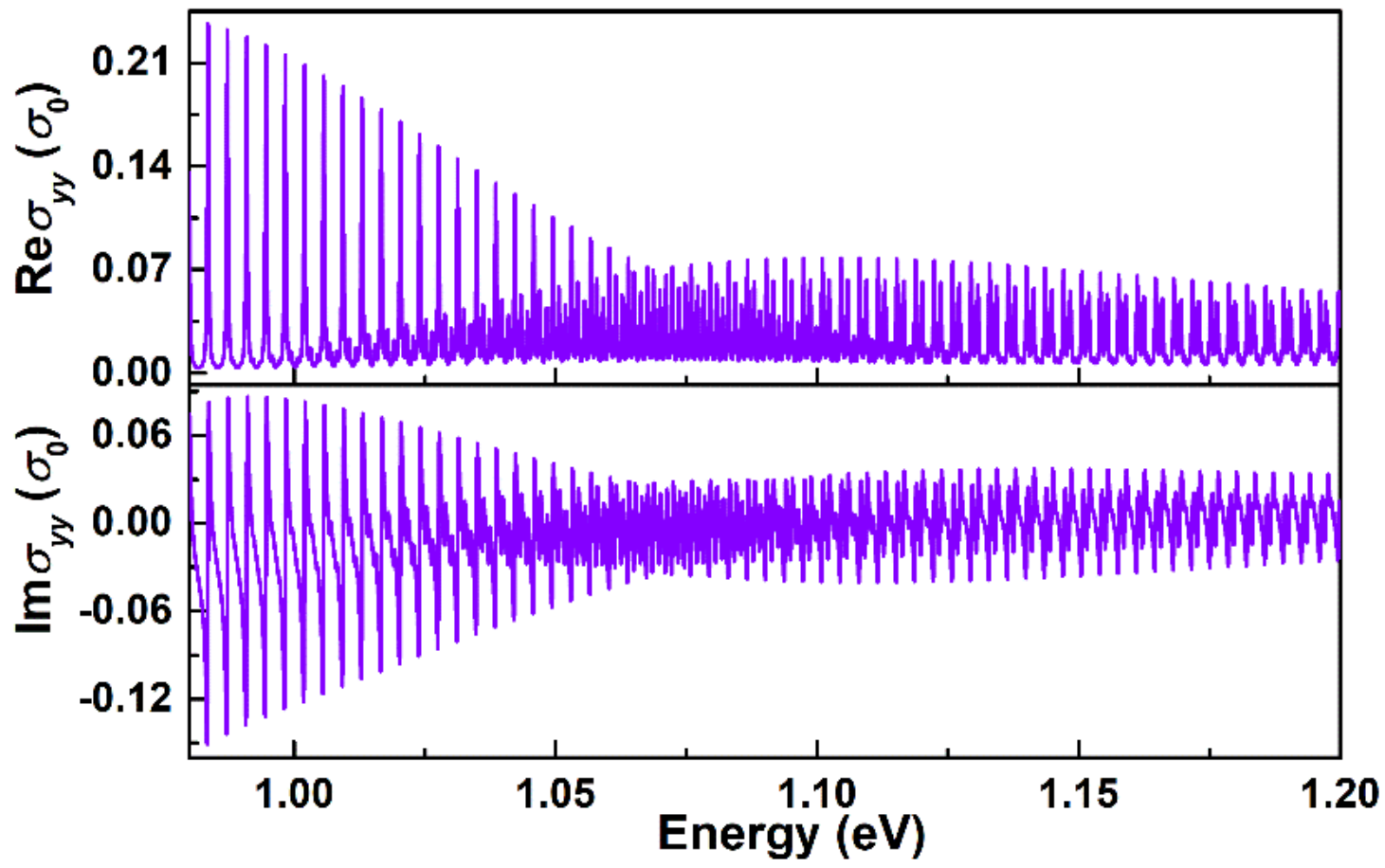

**Fig. S2** Real and imaginary parts of $\sigma_{yy}$ as a function of the photon energy for monolayer $WTe_2$ in a magnetic field with $B$ = 10 T.

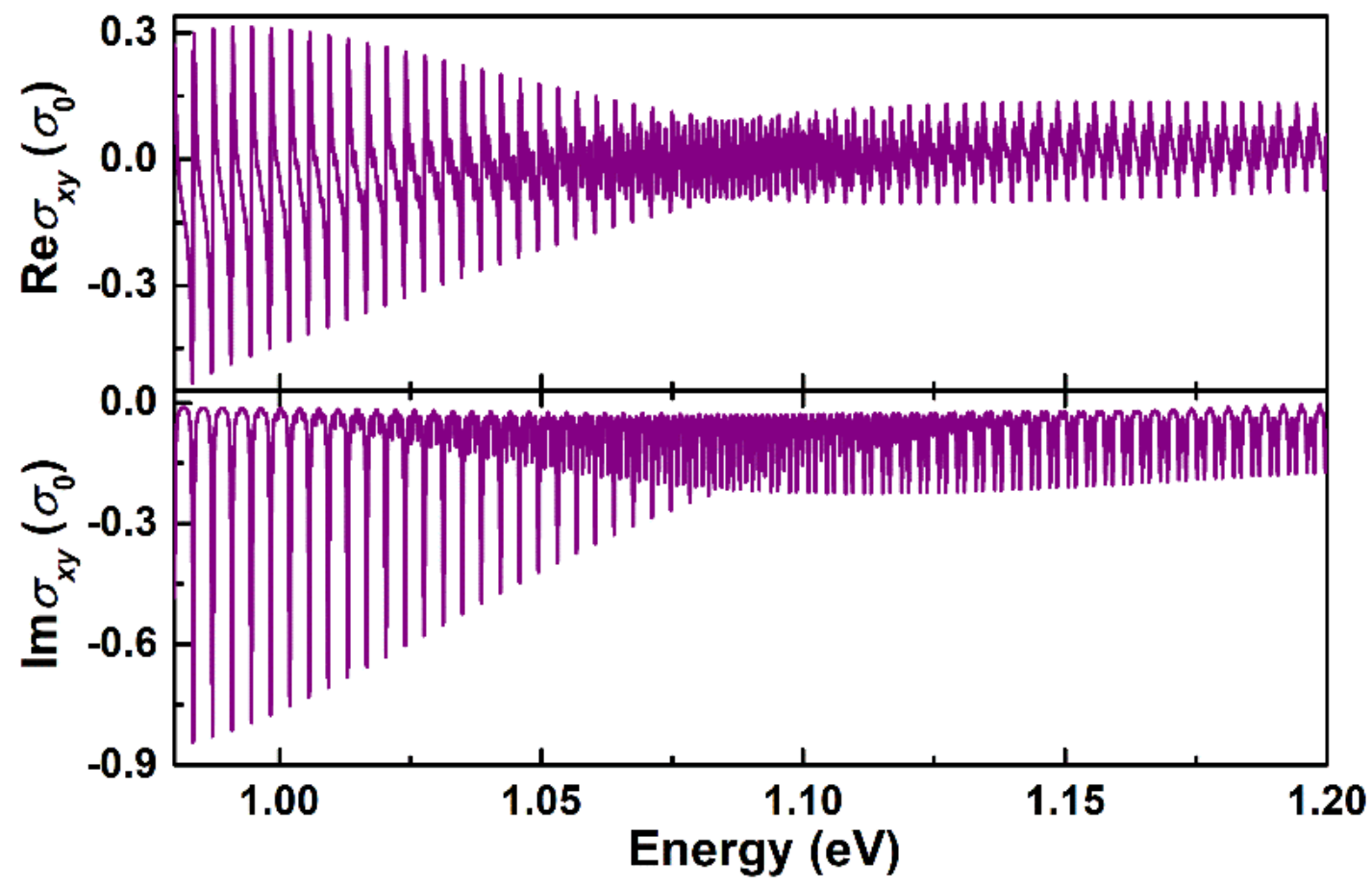

**Fig. S3** Real and imaginary parts of $\sigma_{xy}$ as a function of the photon energy for monolayer $WTe_2$ in a magnetic field with $B$ = 10 T.

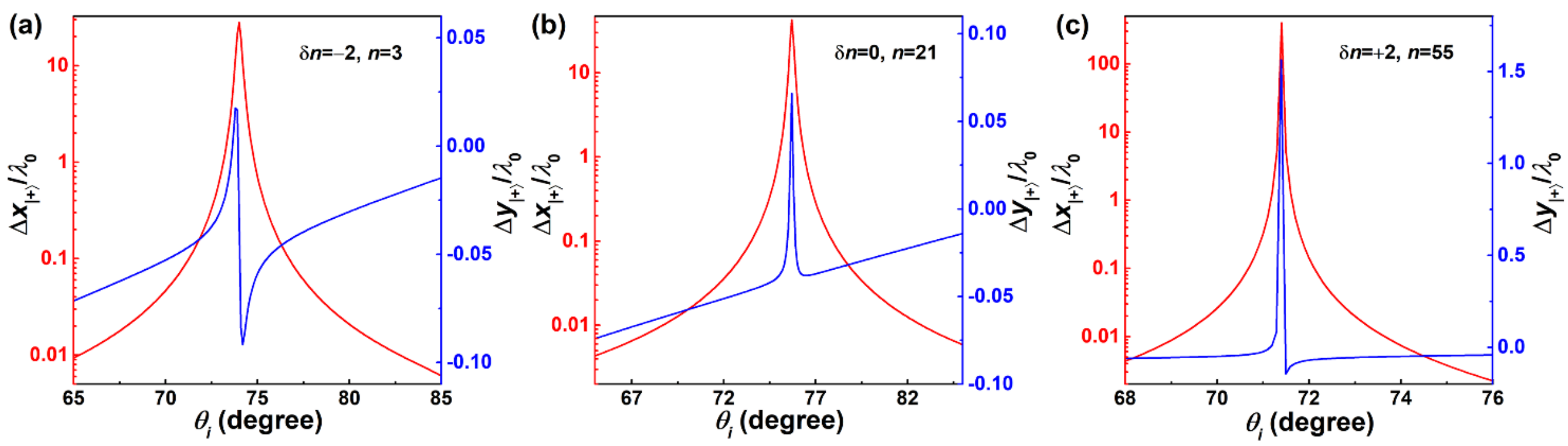

**Fig. S4** In-plane and transverse photonic spin Hall shifts at inter-LL transitions (a) $|n=3\rangle\rightarrow|n'=1\rangle$, (b) $|n=21\rangle\rightarrow|n'=21\rangle$, and (c) $|n=55\rangle\rightarrow|n'=57\rangle$.

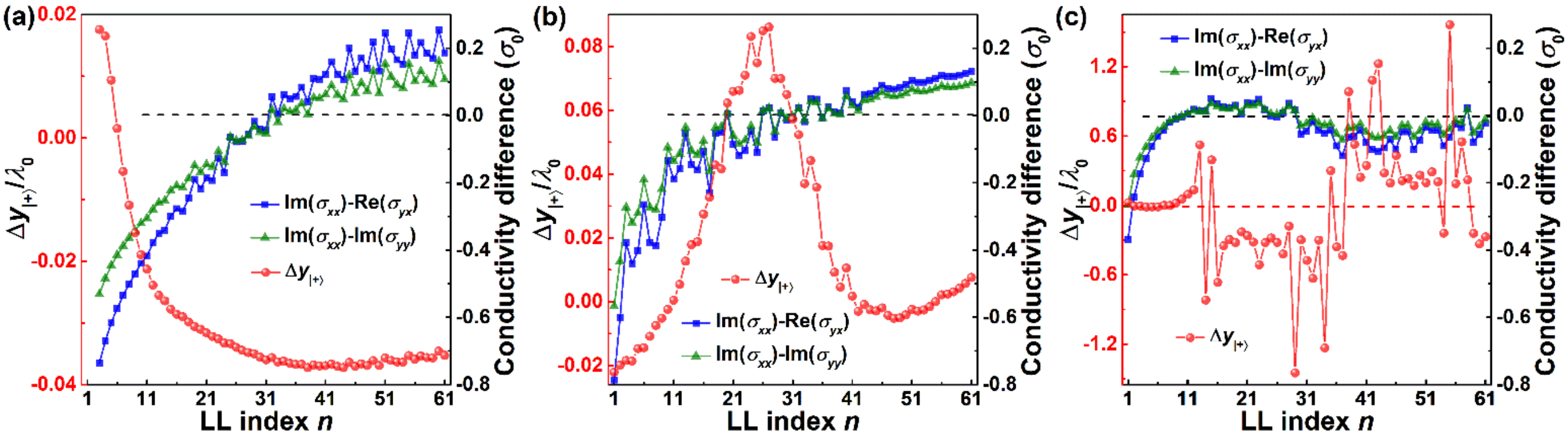

**Fig. S5** The evaluation of extremum displacement of $\Delta y_{|+\rangle}$ with LL index $n$. (a), (b), and (c) are for inter-LL transitions $\delta n$ = −2, 0, and +2, respectively. For comparison, LL-dependent conductivity differences Im($\sigma_{xx}$)−Im($\sigma_{yy}$) and Im($\sigma_{xx}$)−Re($\sigma_{yx}$) are also shown. The graduation lines at which the vertical coordinate axis equals zero are marked by dashed horizontal lines. Symbols are the calculated values while solid lines are just drawn as a guide to the eye. The magnetic field $B$ = 10 T.

## II. Variations of in-plane and transverse spin-dependent shifts with Hall angles

**Table S1** The values of fit coefficients $a_i$, $b_i$, and $c_i$ in Eq. (11) in the main text

| Coefficients | In-plane spin-dependent shift $\Delta x_{\|+\rangle}$ | | | Transverse spin-dependent shift $\Delta y_{\|+\rangle}$ | | |
|---|---|---|---|---|---|---|
| | $\delta n = -2$ | $\delta n = 0$ | $\delta n = +2$ | $\delta n = -2$ | $\delta n = 0$ | $\delta n = +2$ |
| $a_1$ | $3.680\times10^{13}$ | $2.214\times10^{4}$ | 383.9 | -0.02308 | 0.2140 | 1.906 |
| $a_2$ | 54.12 | 183.6 | 113.4 | -0.01131 | 0.1354 | 0.6687 |
| $a_3$ | 1.830 | 3.248 | -69.59 | -0.004125 | -0.2371 | -1.574 |
| $a_4$ | 0.0000 | -0.9525 | -176.3 | -0.03674 | 0.0000 | -0.5130 |
| $a_5$ | 4.679 | $42.86\times10^{13}$ | 0.0000 | 1.084 | 0.007605 | 0.0000 |
| $b_1$ | 1.913 | 0.3668 | 0.1183 | -0.04469 | 0.03268 | 0.1182 |
| $b_2$ | 0.6385 | 0.6328 | 0.08253 | -0.08090 | 0.2918 | 0.08458 |
| $b_3$ | -4.993 | -0.4221 | 0.1732 | -0.1378 | 0.03048 | 0.1497 |
| $b_4$ | 0.8370 | -0.4809 | 0.1538 | -1.938 | -6.156 | 0.1744 |
| $b_5$ | 0.09383 | 159.3 | 2732 | 0.6950 | -6.499 | 0.0000 |
| $c_1$ | 0.3607 | 0.1135 | 0.007005 | 0.01238 | 0.06022 | 0.006291 |
| $c_2$ | 0.6135 | 0.4579 | 0.05332 | 0.009100 | 0.4328 | 0.06808 |
| $c_3$ | 24.74 | 0.2379 | 0.1121 | 0.007374 | 0.08778 | 0.005690 |
| $c_4$ | 0.09004 | 0.05807 | 0.003911 | 1.346 | 0.07238 | 0.1348 |
| $c_5$ | 1.349 | 28.19 | 250.2 | 0.4158 | 1.776 | 250.2 |

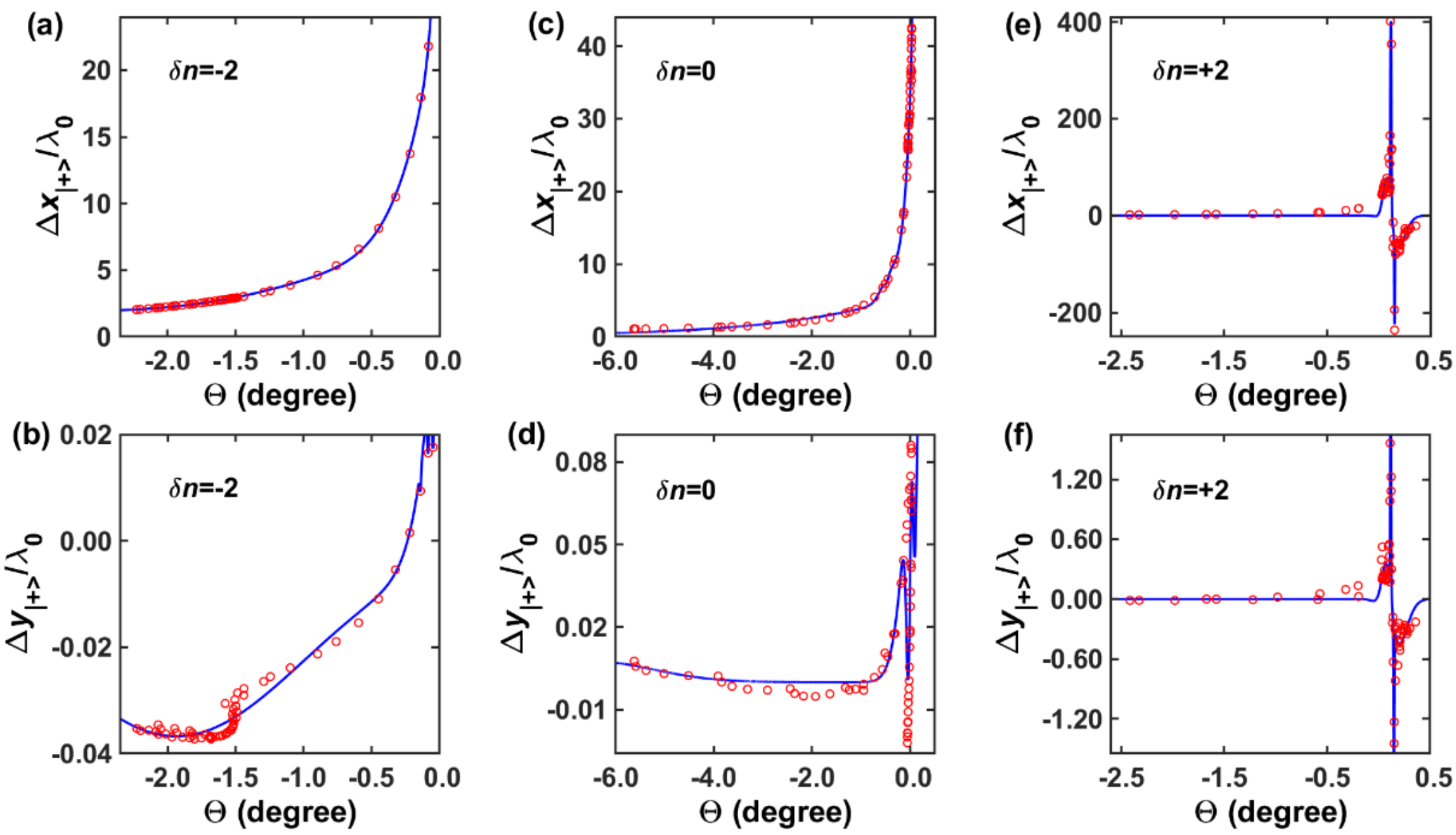

**Fig. S6** Extremum displacements of (a)(c)(e) $\Delta x_{|+\rangle}$ and (b)(d)(f) $\Delta y_{|+\rangle}$ as a function of Hall angle $\Theta$. (a)(b), (c)(d), and (e)(f) are for inter-LL transitions $\delta n = -2$, 0, +2, respectively. Red circles are the results calculated by Eqs. (9) and (10) in the main text. Blue lines are the fitting results by using the empirical Eq. (11) in the main text.

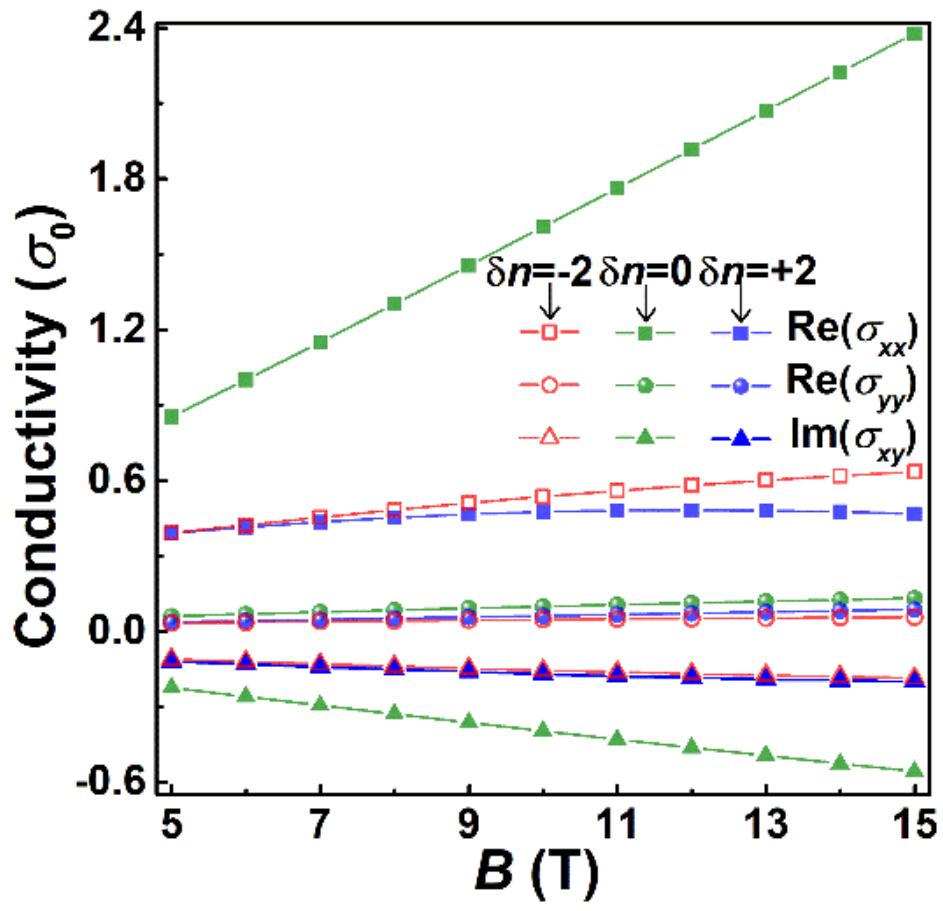

**Fig. S7** At inter-LL transitions $\delta n = 0, \pm 2$ with $n = 21$, optical conductivity components $\mathrm{Re}(\sigma_{xx})$, $\mathrm{Re}(\sigma_{yy})$, and $\mathrm{Im}(\sigma_{xy})$ versus magnetic induction intensity $B$.

## III. PSHE on the surface of monolayer $WTe_2$ without the external magnetic field

Without the external magnetic field, Hall conductivity $\sigma_{xy}$ equals to zero, longitudinal and transverse optical conductivities of monolayer $WTe_2$ can be calculated by [32]

$$\sigma_{jj}(\omega)=\frac{i}{\pi}\frac{D_{jj}}{\omega+i\Gamma}+\frac{i}{\pi}\frac{\omega S_{jj}}{\omega^2-\omega_b^2+i\omega\eta}, \tag{S9}$$

where the index $j = x$ or $y$, $D$ and $S$ denote the spectral weights, $\omega_b$ is the frequency of interband resonance, $\Gamma$ and $\eta$ are the scattering widths for intraband and interband transitions, respectively. According to the experimental results in [32], the fitted parameters are $D_{xx}$=4.49×10$^{11}$ $\Omega^{-1}\cdot s^{-1}$, $D_{yy}$= 8.08×10$^{11}$ $\Omega^{-1}\cdot s^{-1}$, $S_{xx}$=8.07×10$^{11}$ $\Omega^{-1}\cdot s^{-1}$, $S_{yy}$= 4.31×10$^{11}$ $\Omega^{-1}\cdot s^{-1}$, $\omega_b$ = 710 cm$^{-1}$, $\Gamma$= 60 cm$^{-1}$ and $\eta$= 170 cm$^{-1}$. Fig. S8 shows the simulated real and imaginary parts of $\sigma_{xx}$ and $\sigma_{yy}$ of monolayer $WTe_2$ at $B$ = 0 T.

If the $\sigma_{xx}$ axis of $WTe_2$ has a rotation angle $\phi$ to the $x$-$z$ plane, the anisotropy of monolayer atomic film is characterized via an effective conductivity tensor

$$\hat{\sigma}_{eff}=\begin{pmatrix}\cos\phi & \sin(-\phi)\\ \sin\phi & \cos\phi\end{pmatrix}\begin{pmatrix}\sigma_{xx} & 0\\ 0 & \sigma_{yy}\end{pmatrix}\begin{pmatrix}\cos\phi & \sin\phi\\ \sin(-\phi) & \cos\phi\end{pmatrix}=\begin{pmatrix}\sigma_{pp} & \sigma_{ps}\\ \sigma_{sp} & \sigma_{ss}\end{pmatrix}, \tag{S10}$$

where

$$\sigma_{pp} = \sigma_{xx}\cos^2\phi + \sigma_{yy}\sin^2\phi, \tag{S11}$$

$$\sigma_{ps} = \sigma_{sp} = (\sigma_{xx}-\sigma_{yy})\sin\phi\cos\phi, \tag{S12}$$

$$\sigma_{ss} = \sigma_{xx}\sin^2\phi + \sigma_{yy}\cos^2\phi. \tag{S13}$$

The nondiagonal response $\sigma_{ps}$, mixing $p$- and $s$- polarized waves, arises here not due to an intrinsic chirality or magnetism of 2D material, but follows only from a nonzero tilt of the incident plane of light with respect to the main axis, which corresponds to a so-called extrinsic chirality.

Figs. S9(a) and (b) present transverse spin Hall shift $\Delta y_{|+\rangle}$ as a function of the photonic energy and the incident angle at $\phi$ = 0° and 90° ($\Delta x_{|+\rangle}$ = 0 at these two $\phi$ values), respectively. Fig. S9(c) and (d) exhibits in-plane and transverse spin Hall shifts as a function of the photonic energy and the incident angle at $\phi$ = 45°. One can see that all the photonic spin Hall shifts are smaller than 0.15$\lambda_0$. Furthermore, if monolayer $WTe_2$ is substituted by an isotropic 2D material ($\sigma_{xx} = \sigma_{yy}$, $\sigma_{xy} = 0$), in-plane spin Hall shift would disappear and PSHE would be completely independent on the rotation angle $\phi$. These evidence that Hall conductivity $\sigma_{xy}$ induced via magnetization of $WTe_2$ contributes to dramatic enhancement of PSHE in the main text.

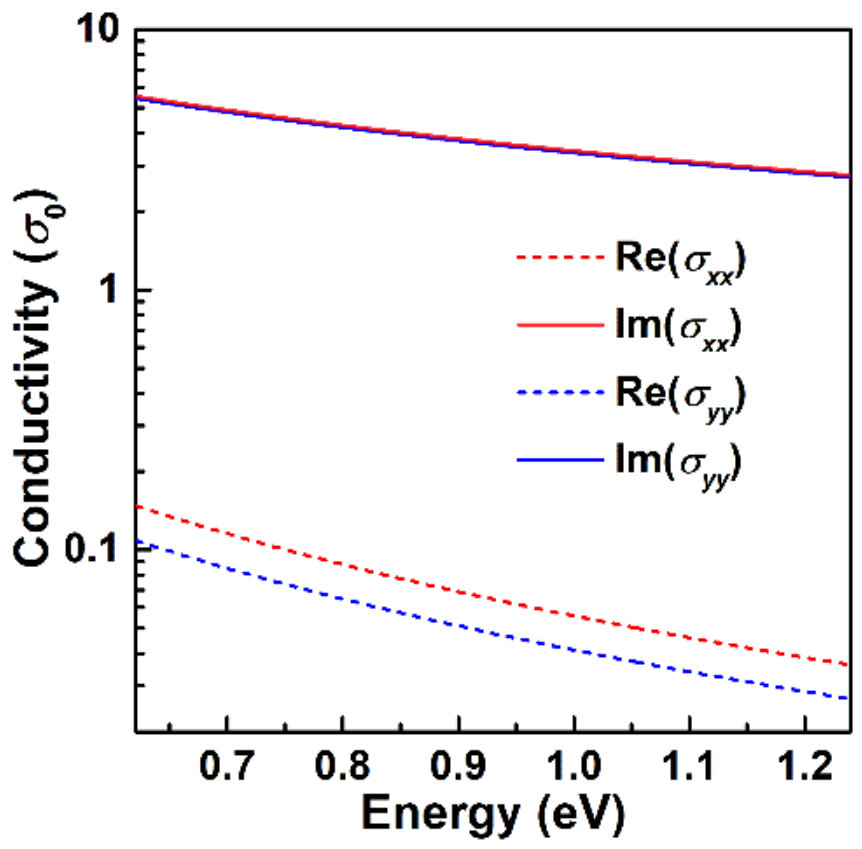

**Fig. S8** Real and imaginary parts of $\sigma_{xx}$ and $\sigma_{yy}$ of monolayer $WTe_2$ versus photon energy at $B$ = 0 T.

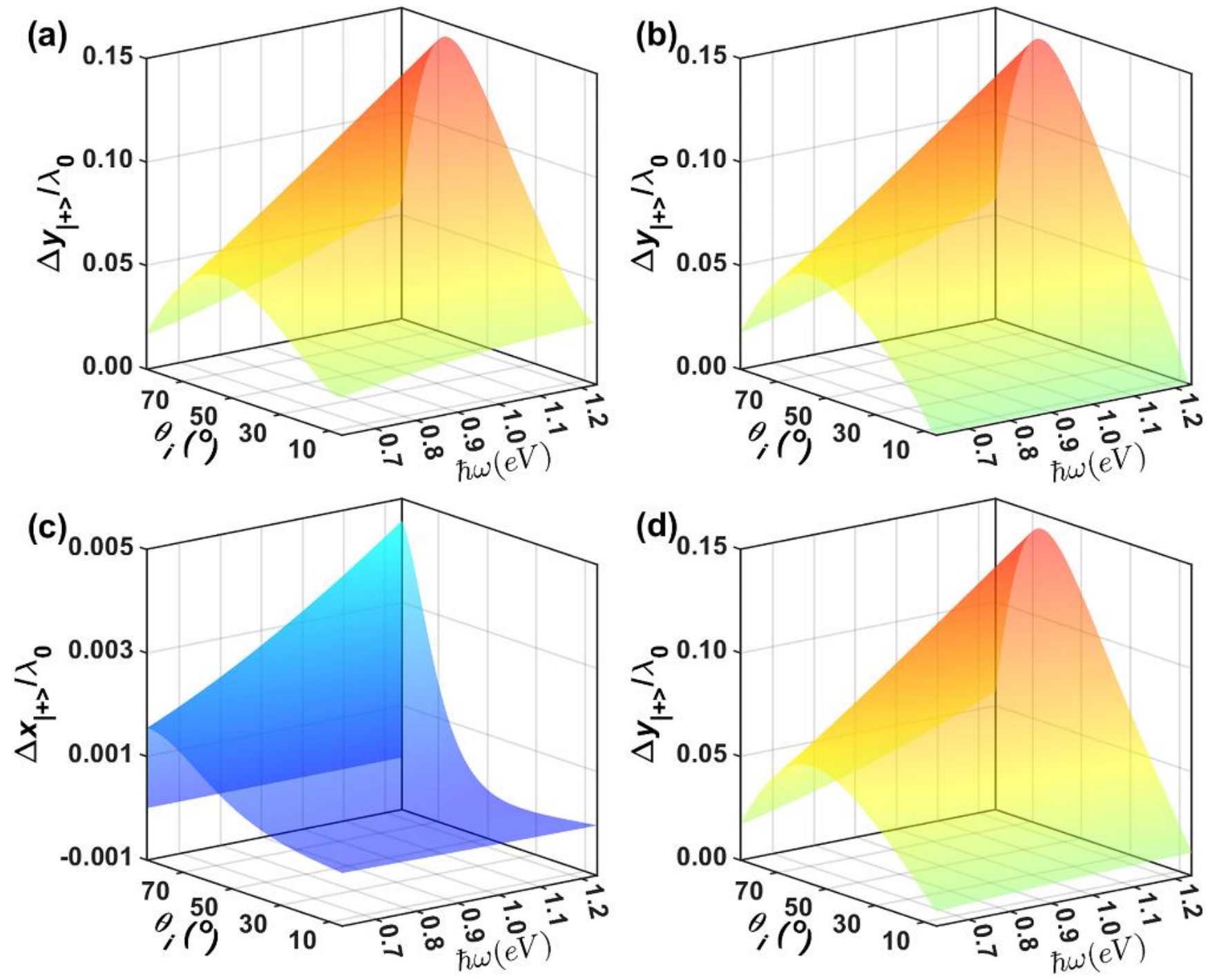

**Fig. S9** At $B$ = 0 T, transverse spin Hall shift $\Delta y_{|+\rangle}$ as a function of the photonic energy and the incident angle at $\phi$ = 0° (a) and $\phi$ = 90° (b). The in-plane spin Hall shift $\Delta x_{|+\rangle}$ is zero at $\phi$ = 0° and 90°. At $B$ = 0 T, (c) in-plane and (d) transverse spin Hall shifts as a function of the photonic energy and the incident angle at $\phi$ = 45°.